\documentstyle[preprint,aps,epsf,floats]{revtex}

\newcommand{\nn}{\nonumber}
\newcommand{\ppp}{\mbox{$({\mathbf p'}-{\mathbf p})^2$}}
\newcommand{\spp}{\mbox{$i\bsigma\cdot ({\mathbf p'\times p})$}}
\newcommand{\two}{{\rm 2}}

\def\xslash#1{{\rlap{$#1$}/}}
\def\Dsl{\hbox{/\kern-.6000em D}} 

\def\dsl{\,\raise.15ex\hbox{/}\mkern-13.5mu D}
\def\bsigma{\mbox{\boldmath $\sigma$}}

\def\psip#1{\psi_{\mathbf{#1}}}
\def\chip#1{\chi_{\mathbf{#1}}}
\def\bsigma{\mbox{\boldmath $\sigma$}}

\def\abs#1{\left| #1 \right|}
\def\ltap{\ \raise.3ex\hbox{$<$\kern-.75em\lower1ex\hbox{$\sim$}}\ }
\def\gtap{\ \raise.3ex\hbox{$>$\kern-.75em\lower1ex\hbox{$\sim$}}\ }
\def\OMIT#1{}

\begin{document}
\setlength\baselineskip{20pt}

\preprint{\tighten  \hbox{UCSD/PTH 99-23}
}

\title{Renormalization group analysis of the QCD quark \\
potential to order $v^2$}

\author{Aneesh V. Manohar\footnote{amanohar@ucsd.edu} and
Iain W.\ Stewart\footnote{iain@schwinger.ucsd.edu} \\[4pt]}
\address{\tighten Department of Physics, University of California at San
Diego,\\[2pt] 9500 Gilman Drive, La Jolla, CA 92099 }

\maketitle

{\tighten
\begin{abstract}

A one-loop renormalization group analysis of the order $v^2$ relativistic
corrections to the static QCD potential is presented. The velocity
renormalization group is used to simultaneously sum $\ln(m/mv)$ and
$\ln(m/mv^2)$ terms. The results are compared to previous calculations in the
literature.

\end{abstract}
}
\vspace{0.7in}

\newpage\tighten

The quark-antiquark interaction potential is needed to compute properties of
heavy quark systems, such as $\Upsilon$ mesons, or $\bar t t$  production near
threshold. In this paper, we will study the renormalization group improved
potential at one-loop. The calculation makes use of NRQCD, formulated as an
effective theory with an expansion in the velocity,
$v$~\cite{Caswell,BBL,Labelle,lm,Manohar,gr,ls,Pineda,P2,Beneke,Gries,P3,kp}. The
leading order term in $v$ is the Coulomb potential, which has been computed to
two-loop order~\cite{Peter,Schroder} using QCD perturbation theory. The
renormalization group running of this term is given by the QCD $\beta$-function,
as is well-known. We will compute the one-loop running of the order $v$ and $v^2$
corrections to the quark potential, using a formulation of NRQCD introduced
recently~\cite{LMR}, and assuming $mv^2\gg \Lambda_{\rm QCD}$.  In QCD the
one-loop potential to order $v^2$ has been computed
previously~\cite{Gupta,Yndurain}. For $\mu=m$ the logarithms in the effective
theory must agree with the logarithmic terms in these computations.  We find
agreement when some previously neglected terms are included in the
spin-independent part of the quark potential. The renormalization group analysis
allows one to resum logarithms of $v$ in the effective theory, which is not
possible using only the one-loop quark potential.

The formalism we will use has been described in Ref.~\cite{LMR}, and will be
called vNRQCD.\footnote{Following a suggestion of A.H. Hoang.}  In vNRQCD, one
matches onto QCD at $\mu=m$ and then runs to lower scales in the effective
theory using a velocity renormalization group (VRG). The VRG allows one to
simultaneously sum logarithms of $mv$ and $mv^2$ in the effective theory. In an
alternative approach, called pNRQCD~\cite{Pineda}, the matching takes place in
two stages, at $\mu=m$ and then at $\mu=m v$. The logarithmic corrections to
the potential were recently computed using pNRQCD by Brambilla et
al.~\cite{Brambilla}.  Our results agree with theirs when the resummed
logarithms are expanded to linear order.

The vNRQCD effective Lagrangian is written in terms of fields $\psip p$ which
annihilate a quark, $\chip p$ which annihilate an antiquark, $A^\mu_{p}$ which
annihilate and create soft gluons, and $A^\mu$ which annihilate and create
ultrasoft gluons. The covariant derivative is $D^\mu =
\partial^\mu + i g A^\mu=(D^0,-\mathbf{D})$, so that $D^0=\partial^0+ig A^0$,
${\mathbf D}={\mathbf \nabla}-ig{\mathbf A}$, and involves only the ultrasoft
gluon fields. The ultrasoft piece of the effective Lagrangian we need contains
the quark, antiquark, and ultrasoft gluon kinetic energies,
\begin{eqnarray} \label{Lu}
  {\mathcal L}_u &=& -{1\over 4}F^{\mu\nu}F_{\mu \nu} + \sum_{\mathbf p}
   \psip p ^\dagger   \Biggl\{ i D^0 - {\left({\bf p}-i{\bf D}\right)^2
   \over 2 m} +\frac{{\mathbf p}^4}{8m^3} \Biggr\} \psip p  \nn\\
   & & + \chip p ^\dagger
   \Biggl\{ i D^0 - {\left({\bf p}-i{\bf D}\right)^2
   \over 2 m} +\frac{{\mathbf p}^4}{8m^3} \Biggr\} \chip p \,,
\end{eqnarray}
where the covariant derivative on $\psip p$ and $\chip p$ contain the color
matrices $T^A$ and $\bar T^A$ for the $\bf 3$ and $\bf {\bar 3}$
representations, respectively. The coefficients in Eq.~(\ref{Lu}) do not
run due to reparameterization invariance \cite{repar}. The potential
interaction is
\begin{eqnarray}\label{1}
{\mathcal L}_p= - V_{\alpha\beta\lambda\tau} \left({\bf p},{\bf
  p^\prime}\right)\ {\psip {p^\prime}}_\alpha^\dagger\: {\psip p}_\beta\:
  {\chip {-p^\prime}}_\lambda^\dagger\:  {\chip {-p}}{}_\tau,
\end{eqnarray}
where $\alpha,\beta,\lambda,\tau$ denote color and spin indices. It is
convenient to write the terms in $V$ in matrix form. For example,
\begin{eqnarray} \label{pCoul}
V = {4 \pi \alpha_s \over \left( {\mathbf p - p^\prime} \right)^2}\:
  (T^A \otimes \bar T^A)
\end{eqnarray}
represents the Coulomb interaction, and corresponds to Eq.~(\ref{1}) with
\begin{eqnarray}
V_{\alpha\beta\lambda\tau}\left({\bf p},{\bf p^\prime}\right) =
  {4 \pi \alpha_s \over \left( {\mathbf p - p^\prime} \right)^2}\
  T^A_{\alpha\beta}\,  \bar T^A_{\lambda\tau} \,.
\end{eqnarray}
\begin{figure}
  \epsfxsize=10cm \hfil\epsfbox{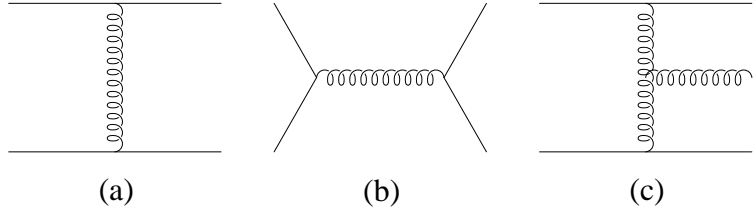}\hfill
{\tighten \caption{QCD diagrams for tree level matching.} \label{fig_tree} }
\end{figure}
The diagrams in Fig.~\ref{fig_tree}a,b give terms of the form
\begin{eqnarray}
 V &=&  (T^A \otimes \bar T^A) \left[
 { {\cal V}_c^{(T)} \over {\mathbf k}^2}
 + { {\cal V}_r^{(T)}\: ({\mathbf p^2 + p^{\prime 2}}) \over 2\, m^2\,
 {\mathbf k}^2}
 + { {\cal V}_s^{(T)} \over m^2}\, {\mathbf S}^2 + {{\cal V}_\Lambda^{(T)}
 \over m^2}\, \Lambda({\mathbf p^\prime ,p}) + { {\cal V}_t^{(T)} \over  m^2}\,
 T({\mathbf k})\right] \nn \\
 && + (1\otimes 1)\ { {\cal V}_s^{(1)} \over m^2}\: {\mathbf S}^2 \,,
\end{eqnarray}
where ${\bf k} = {\bf p'} - {\bf p}$ and the notation of Titard and
Yndurain~\cite{Yndurain} has been used:
\begin{eqnarray}
\mathbf S &=& { {\mathbf \bsigma_1 + \bsigma_2} \over 2},
 \qquad \Lambda({\mathbf p^\prime, p }) = -i {\mathbf S \cdot ( p^\prime
 \times p) \over  {\mathbf k}^2 },\qquad
 T({\mathbf k}) = {\mathbf \bsigma_1 \cdot \bsigma_2} - {3\, {\mathbf k
 \cdot \bsigma_1}\,  {\mathbf k \cdot \bsigma_2} \over {\mathbf k}^2} \,.
\end{eqnarray}
The terms with coefficients ${\cal V}_r^{(T)}$, ${\cal V}_s^{(T,1)}$, ${\cal
V}^{(T)}_\Lambda$, and ${\cal V}^{(T)}_t$ are order $v^2$ corrections to the
lowest order Coulomb term, ${\cal V}_c^{(T)}$.  Matching to the tree level
diagram in Fig.~\ref{fig_tree}a at $\mu=m$ gives
\begin{eqnarray}  \label{bc}
 {\cal V}_c^{(T)} &=& 4 \pi \alpha_s(m)\,, \qquad
 {\cal V}_r^{(T)} = 4 \pi \alpha_s(m)\,, \qquad
 {\cal V}_s^{(T)} = -\frac{4 \pi \alpha_s(m)}{3}\,, \nn  \\
 {\cal V}_\Lambda^{(T)} &=& -6 \pi \alpha_s(m) \,,\qquad
 {\cal V}_t^{(T)} = -\frac{\pi \alpha_s(m)}{3} \,, \qquad
 {\cal V}_s^{(1)} =0 \,.
\end{eqnarray}
The annihilation diagram in Fig.~\ref{fig_tree}b gives the additional
contributions
\begin{eqnarray}  \label{bc2}
  {\cal V}_{s,a}^{(T)} =  {1 \over N_c}\: \pi\, \alpha_s(m) \,,\qquad
  {\cal V}_{s,a}^{(1)} =  {(N_c^2-1)\over 2N_c^2}\: \pi\, \alpha_s(m) \,.
\end{eqnarray}
These contributions have been separated from those in Eq.~(\ref{bc}) to
facilitate comparison with results in the literature. In the color singlet
channel there is no annihilation contribution.

We have chosen to use the basis in which the potential $V$ is written as a
linear combination of $1 \otimes 1$ and $T \otimes \bar T$. One can convert to
the singlet and octet potential, using the linear transformation
\begin{eqnarray}
 \left[\begin{array}{c} V_{\rm singlet} \cr V_{\rm octet} \end{array}\right]
 =\left[\begin{array}{ccc} 1 &  & -C_F \cr
    1 &  & {1\over 2} C_A - C_F \cr
 \end{array}\right]
 \left[\begin{array}{c} V_{1\otimes 1} \cr V_{T\otimes T} \end{array}\right]\,,
\end{eqnarray}
where $C_F=(N_c^2-1)/(2N_c)$ and $C_A=N_c$.

Potential terms can also be induced by operator mixing in the renormalization
group flow. For instance, we will see that the running at one-loop induces
order $v^2$ terms of the form
\begin{eqnarray}
 V = (T^A \otimes \bar T^A)\  { {\cal V}_2^{(T)} \over m^2 }
 + ({1} \otimes {1}) \ { {\cal V}_2^{(1)} \over m^2 } \,,
\end{eqnarray}
which are absent at tree level. Additional terms also occur when matching the
potential at higher orders in the loop expansion.  For instance, at one loop
the order $v$ correction
\begin{eqnarray} \label{Lk}
 V_{\rm singlet} = { {\cal V}_k^{(s)}\: \pi^2 \over m\,
 |{\mathbf k}| }
\end{eqnarray}
is generated. Ref.~\cite{Yndurain} has
${\cal V}_k^{(s)}=-C_F (C_A-C_F/2)\, \alpha_s(m)^2$ for the matching at $\mu=m$.

There are terms in the Lagrangian where ultrasoft gluons couple to
potential operators.  For instance, the diagram in Fig.~\ref{fig_tree}c
generates the term
\begin{eqnarray}  \label{Lpu}
 {\cal L}_{pu} = \frac{2ig^2}{({\mathbf p'}-{\mathbf p})^4} f^{ABC}\:
  ({\mathbf p-p'}) \cdot (g{\mathbf A}^C) [{\psip {p^\prime}}^\dagger\:
  T^A {\psip p}\:{\chip {-p^\prime}}^\dagger\: \bar T^B {\chip {-p}}{} ] \:
  \,.
\end{eqnarray}

\begin{figure}
  \epsfxsize=12cm  \hfil\epsfbox{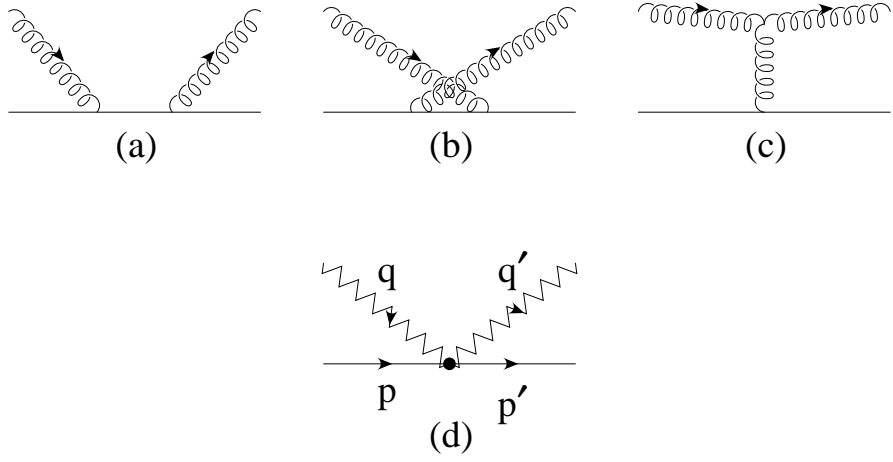} \hfill
{\tighten \caption{The Compton scattering diagrams (a,b,c) generate the soft
gluon coupling (d) in the effective theory. Diagrams analogous to (c) but with
external ghosts or massless quarks generate the soft ghost and massless soft
quark couplings in Eq.~(\ref{Lsoft}). \label{fig_soft}} }
\end{figure}

The terms in the soft Lagrangian that we need for our computation are
\begin{eqnarray}\label{Lsoft}
 {\mathcal L}_s &=& \sum_{q} \bigg\{ \abs{q^\mu A^\nu_q -
 q^\nu A^\mu_q}^2 + \bar \varphi_q\, \xslash{q}\, \varphi_q  +
 \bar c_q \, q^2 c_q \ \bigg\}  \\
 && - g^2 \sum_{{\mathbf p,p'},q,q'} \bigg\{ \frac12\, {\psip
 {p^\prime}}^\dagger\:
 [A^\mu_{q'},A^\nu_{q}] U_{\mu\nu}^{(\sigma)}\: {\psip p}\: + \frac12\,
 {\psip {p^\prime}}^\dagger\: \{A^\mu_{q'},A^\nu_{q}\} W_{\mu\nu}^{(\sigma)}\:
 {\psip p}\: \nn \\
 && + {\psip {p^\prime}}^\dagger\: [\bar c_{q'},c_{q}] Y^{(\sigma)}\:
 {\psip p}\: + ({\psip {p^\prime}}^\dagger\: T^B Z_\mu^{(\sigma)}\:
 {\psip p} ) \:(\bar \varphi_{q'} \gamma^\mu T^B \varphi_q)  \bigg\}
 + (\psi \to \chi,\: T\to \bar T) \,. \nn
\end{eqnarray}
Here $U$, $W$, $Y$, and $Z$ are functions of $({\bf p},{\bf p^\prime},q,q')$
and matrices in spin and are generated by integrating out the intermediate
offshell quarks and gluons in Fig.~\ref{fig_soft}. The field $c_{q}$ is the
soft ghost field, and $\varphi_q$ is the massless soft quark field with $n_f$
flavor components. The index $\sigma$ denotes the relative order in the $v$
expansion.
Performing the matching in Fig.~\ref{fig_soft} in Feynman gauge we find
\begin{eqnarray}  \label{nlsoft}
 U^{(0)}_{00} &=&  \frac{1}{q^0}\,,\quad
 U^{(0)}_{0i}  = -\frac{{\mathbf (\two p'-\two p-q)}^i}{\ppp}\,, \quad
 U^{(0)}_{i0}  = -\frac{{\mathbf (p-p'-q)}^i}{\ppp}\,, \quad
 U^{(0)}_{ij}  = \frac{-2 q^0 \delta^{ij} }{\ppp} \nn \,,\\[5pt]
 U^{(1)}_{00} &=& \frac{ {\mathbf (p'+p)\cdot q}}{2m (q^0)^2} -
    \frac{ {\mathbf (p'+p)\cdot q}}{m \ppp} -\frac{i c_F\, \bsigma\cdot
    [{\mathbf q\times (p-p')}] }{m \ppp} \,, \\[5pt]
 U^{(1)}_{0i} &=& -\frac{{\mathbf (p+p')}^i}{2m q^0} + \frac{i c_F
   ({\mathbf q} \times\bsigma)^i}{2 m q^0} + \frac{q^0 {\mathbf (p+p')}^i}
   {2m \ppp} + \frac{i c_F\, q^0 [{\mathbf (p-p')\times\bsigma}]^i}{2m\ppp} \,,
   \nn \\[5pt]
 U^{(1)}_{i0} &=& -\frac{{\mathbf (p+p')}^i}{2m q^0} - \frac{i c_F [ { \mathbf
  (p-p'+q)\times \bsigma }]^i }{2 m q^0} + \frac{q^0{\mathbf (p+p')}^i}{2m \ppp}
  + \frac{i c_F\, q^0 {[\mathbf (p-p')\times \bsigma}]^i}{2m\ppp} \,,
  \nn \\[5pt]
 U^{(1)}_{ij} &=& \frac{i c_F\, \epsilon^{ijk}\bsigma^k}{2m} + [ 2 \delta^{ij}
  {\mathbf q}^m \!+\! \delta^{im} ( {\mathbf \two p'\!-\! \two p\!-\!q)}^j
  \!+\! \delta^{jm} {\mathbf (p\!-\!p'\!-\!q)}^i]  \nn\\*
 && \times   \Big[ \frac{{\mathbf
  (p\!+\!p')}^m\!+\! i c_F\, \epsilon^{mkl} {\mathbf (p\!-\!p')}^k \bsigma^l }
  {2m ({\mathbf p'-p})^2}
  \Big] \,, \nn\\[5pt]
 U^{(2)}_{00} &=& -\frac{c_D \ppp}{8 m^2 q^0} + \frac{c_S\,\spp}{4 m^2 q^0} +
  \frac{({\bf p\cdot q})^2+({\bf p'\cdot q})^2}{2m^2(q^0)^3}+\frac{(2-c_D)
  {\mathbf (p-p')\cdot q}}{4m^2 q^0} \nn\\
  && +\frac{ (1-c_D) {\mathbf q}^2}{4 m^2 q^0} \,,\nn \\[5pt]
 U^{(2)}_{0i} &=& - \frac{[ {\mathbf p\cdot q}\,{\mathbf (\two p+q)}^i
  +{\mathbf p'\cdot q}\: {\mathbf (\two p'-q)}^i\, ]}{ 4 m^2 (q^0)^2 }
  + \frac{ ic_F [{\mathbf q \times\bsigma}]^i\:
  {\mathbf (p+p')\cdot q} }{ 4 m^2 (q^0)^2 } \nn\\[5pt]
  && \ +\frac{(c_D-1)({\mathbf p-p'+q})^i}{4 m^2} + \frac{ {\mathbf
   (\two p'-\two p-q)}^i }{ 4 m^2 } \bigg[ \frac{c_D}{2}
   -  \frac{c_S\, \spp}{\ppp} \bigg] \,, \nn\\[5pt]
 U^{(2)}_{i0} &=& - \frac{ [{\mathbf p\cdot q}\,{\mathbf (p+p'+q)}^i +
  {\mathbf p'\cdot q}\: {\mathbf (p+p'-q)}^i\, ] }{ 4 m^2 (q^0)^2 }
  - \frac{ ic_F [{\mathbf (p-p'+q)\times  \bsigma}]^i\:
  {\mathbf (p+p')\cdot q} }{ 4 m^2 (q^0)^2 }   \nn \\[5pt]
  && \ +\frac{(c_D-1) {\mathbf q}^i}{4 m^2} + \frac{ {\mathbf (p-p'-q)}^i }
  { 4 m^2 } \bigg[ \frac{c_D}{2} - \frac{c_S\, \spp}{\ppp} \bigg] \,, \nn\\[5pt]
 U^{(2)}_{ij} &=& \frac{{\mathbf (p+p')}^i {\mathbf (p+p')}^j}{4 m^2 q^0} +
  \frac{ c_F^2 {\mathbf (p-p') \cdot q}\: \delta^{ij}}{4 m^2 q^0}
  + \frac{ i c_F {\mathbf (p+p')}^j\: [ {\mathbf (p-p')
  \times \bsigma }]^i }{ 4 m^2 q^0 } \nn\\[5pt]
  && \ - \frac{ i c_F\, \epsilon^{ijk} {\mathbf q}^k
  {\mathbf \bsigma \cdot (p+p') }}{4 m^2 q^0} +  \frac{ i c_F\, \epsilon^{ijk}
  \bsigma^k {\mathbf (p+p')\cdot q }}
  {4 m^2 q^0} + \frac{(1-c_F^2) {\mathbf q}^i({\mathbf p-p'+q})^j}{4 m^2 q^0}
  \nn\\[5pt]
 && \ + \frac{ (c_F^2-c_D) {\mathbf q}^2 \delta^{ij} }{4 m^2 q^0}
  + \frac{ \delta^{ij} q^0}{2 m^2 } \bigg[ \frac{c_D}{2}
  -\frac{c_S\, \spp}{\ppp} \bigg] \,, \nn \\
 W^{(0)}_{\mu\nu} &=& 0 \,,\nn\\
 W^{(1)}_{00} &=& \frac{1}{2m} +\frac{{\mathbf (p-p')\cdot q}}{2m (q^0)^2}
   \,,\quad
 W^{(1)}_{0i}  = -\frac{{\mathbf (p-p'+q)}^i}{2m q^0} \,,\quad
 W^{(1)}_{i0}  = \frac{-{\mathbf q}^i}{2m q^0} \,,\quad
 W^{(1)}_{ij}  = \frac{\delta^{ij}}{2m} \,, \nn \\[5pt]
 Y^{(0)} &=&   \frac{-q^0}{\ppp} \,,\quad
 Y^{(1)} = \frac{ {\mathbf q \cdot (p+p')} + ic_F\, {\mathbf \bsigma \cdot
    [q\times (p-p')] } }{2m \ppp }  \,, \quad \nn\\[5pt]
 Y^{(2)} &=& \frac{ c_D\,q^0 }{8m^2} - \frac{c_S\, \spp q^0 }{4 m^2 \ppp}
    \,,\nn \\
 Z^{(0)}_0 &=& \frac{1}{\ppp} \,,\quad  Z^{(0)}_i =0 \,, \quad
 Z^{(1)}_0 = 0 \,, \quad
 Z^{(1)}_i =  \frac{ -({\mathbf p+p'})^i - i c_F [({\mathbf p-p')\times
 \bsigma]}^i}
   {2m \ppp } \,,\quad \nn\\
 Z^{(2)}_0 &=& -\frac{1}{4 m^2} \bigg[ \frac{c_D}{2} - \frac{c_S\, \spp}
   {\ppp} \bigg] \,,\quad
 Z^{(2)}_i = 0 \,.  \nn
\end{eqnarray}
$W^{(2)}_{\mu\nu}$ is not required for the calculation here. In
Eq.~(\ref{nlsoft}) we have set ${\mathbf p^2}={\mathbf p'}^2$, since this case
is sufficient for our analysis. One could also set $q^0=|{\mathbf q}|$, however
for calculational purposes we found it simpler to keep factors of $q^0$
explicit and ignore the energy poles this generates on the real axis.

The coefficients of operators in the soft effective Lagrangian can run.  In the
soft regime the offshell quark propagators are identical to the quark
propagators in heavy quark effective theory (HQET) \cite{Beneke,Gries}.
Therefore, the divergence structure in the soft regime is the same as HQET. The
running of ${\cal L}_s$ can be computed indirectly using known results for
the HQET Lagrangian,
\begin{eqnarray}
  {\cal L}_{\rm HQET} = \psi^\dagger \bigg\{ i D^0 +
    { {\mathbf D}^2 \over 2 m} +
    \frac{c_F g}{2m} {\mathbf \bsigma \cdot B} + \
    \frac{c_D g}{8 m^2} [ {\mathbf D \cdot E} ] +
    \frac{i c_S g}{8 m^2} \bsigma \cdot ( {\mathbf D \times E-E\times D})
   \bigg\} \psi \,.
\end{eqnarray}
The soft Lagrangian at $\mu=m\nu$ can be computed by first scaling the HQET
Lagrangian to $\mu=m\nu$, and then matching to the soft Lagrangian by computing
the Compton scattering amplitude using HQET vertices in 
Fig.~\ref{fig_soft}a,b,c. This gives the $c_F$, $c_D$ and $c_S$ dependence in
Eq.~(\ref{nlsoft}). The running coefficients $c_F$, $c_D$, and $c_S$ were
computed in  Refs.~\cite{run1,run2,run3,run4}:
\begin{eqnarray}
   c_F(\nu) &=& z^{-C_A/\beta_0}\,, \qquad  c_S(\nu) = 2 z^{-C_A/\beta_0} -1
    \,,\\[5pt]
  c_D(\nu) &=& z^{-2C_A/\beta_0} +\Big(\frac{20}{13}+\frac{32 C_F}{13 C_A}\Big)
  \bigg[ 1 - z^{-13C_A/(6\beta_0)} \bigg] \nn \,.
\end{eqnarray}
Here $z = { \alpha_s(m\nu)/ \alpha(m) }$ and $\beta_0 = {11 C_A/3-4 T_F n_f /
3}$, where $T_F=1/2$ is the index of the fermion representation.

The renormalization group equation for the potential is
\begin{eqnarray}
&&\mu {d \over d \mu} \left[\begin{array}{c} V_{1\otimes 1} \cr V_{T\otimes T}
\end{array}\right] = { \alpha_{s} \over \pi}  \: \Gamma \:
\left[\begin{array}{c} V_{1\otimes 1} \cr V_{T\otimes T} \end{array}\right] \,,
\label{3}
\end{eqnarray}
where $\Gamma$ is a $2\times 2$ matrix and can be calculated as a power series
in $v$ and $\alpha_s$.  The one-loop ultrasoft contributions to $\Gamma$ are
straightforward to compute. Assume that the potential has the form
\begin{eqnarray} \label{Xptnl}
   V=\left(X^A \otimes \bar X^A\right)\ V\left({\bf p},{\bf p^\prime}\right),
\end{eqnarray}
where $X^A$ is either $T^A$ or 1.  An ultrasoft loop integrates over the
ultrasoft loop-momentum, and due to the multipole expansion does not involve
the labels ${\bf p}$ and ${\bf p^\prime}$, so one can compute all one-loop
divergent graphs with a single insertion of $V$. In Feynman gauge, the leading
order graphs involve the ultrasoft vertex from the $D^0$ term in ${\cal L}_u$.
The sum of all the graphs has no ultraviolet divergence, so there is no order
$v^0$ ultrasoft contribution to $\Gamma$~\cite{LMR}.  Thus, the running of the
quark potential involves mixing between different powers of $v$, i.e. running
of the $v^2$ term proportional to the $v^0$ term. To the order we are working,
we need the ultrasoft vertices from the ${\mathbf p \cdot A}/m$ operator, and
insertions of ${\mathbf p \cdot \nabla}/m$.  Graphs with one insertion of
$\nabla^2/(2m)$ or ${\mathbf p}^4/(8m^3)$ are the same order in the power
counting but do not have ultraviolet log divergences. Graphs with one insertion
of the operator in Eq.~(\ref{Lpu}) and one ${\mathbf p \cdot A}/m$ vertex also
do not contribute for this reason. The graphs which contribute to this mixing
are listed in Table~\ref{tab:us} and give
\begin{table} {\tighten
\caption{Ultrasoft contributions to the running of the potential in Feynman
gauge. The dot denotes an insertion of the potential in Eq.~(\ref{Xptnl}). The
$\mathbf p\cdot A$ column gives the momentum factor from diagrams where the
gluon coupling is due to the $\mathbf p\cdot A$ interaction in the Lagrangian.
The $\mathbf p \cdot \nabla $ column gives the factor from graphs in which the
gluon vertices are from the $D^0$ interaction, and there are two insertions of
the $\mathbf p\cdot \nabla $ operator on the quark lines in the loop. In
$d=4-2\epsilon$ dimensions the ultraviolet divergent part of a diagram is $-i
V({\mathbf p,p'}) \alpha_s/(2\pi \epsilon\, m^2)$ times the color and momentum
factors.  The sum of diagrams is gauge independent. \label{tab:us}} }
\[
\begin{array}{lccccc}
\hbox{Diagram} & & \hbox{Color Factor} & {\mathbf p\cdot A} &
 {\mathbf p\cdot \nabla }  \\ \hline\\
\epsfxsize=2cm \lower15pt\hbox{\epsfbox{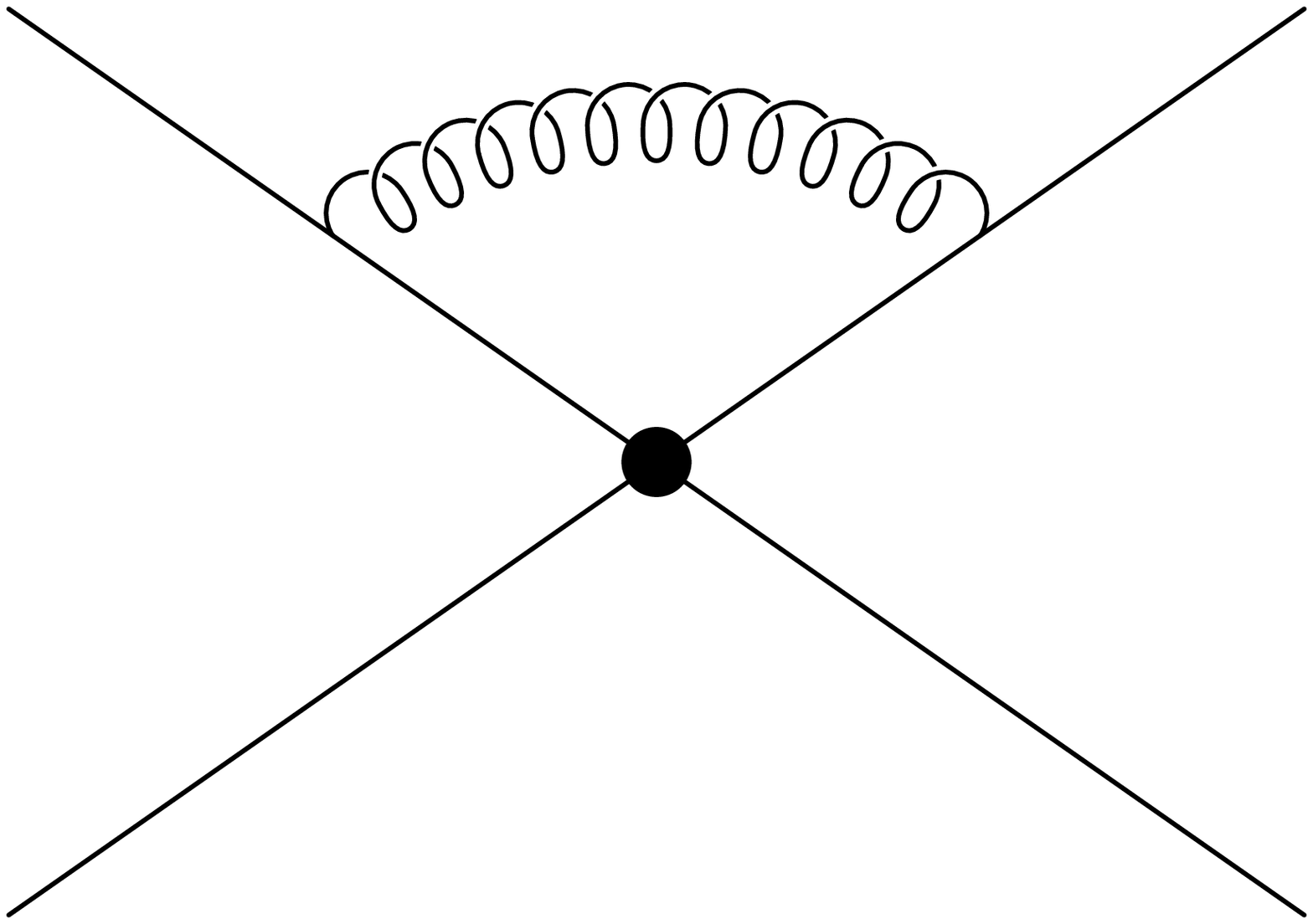}} & \phantom{xxx} &
 T^b X^a T^b \otimes \bar X^a  & {\mathbf p \cdot p^\prime} &
 -{1\over 3}\left({\mathbf p^2}+{\mathbf p^{\prime\,2}}+
 {\mathbf p \cdot p^\prime}\right)  \\ \\
\epsfxsize=2cm \lower15pt\hbox{\epsfbox{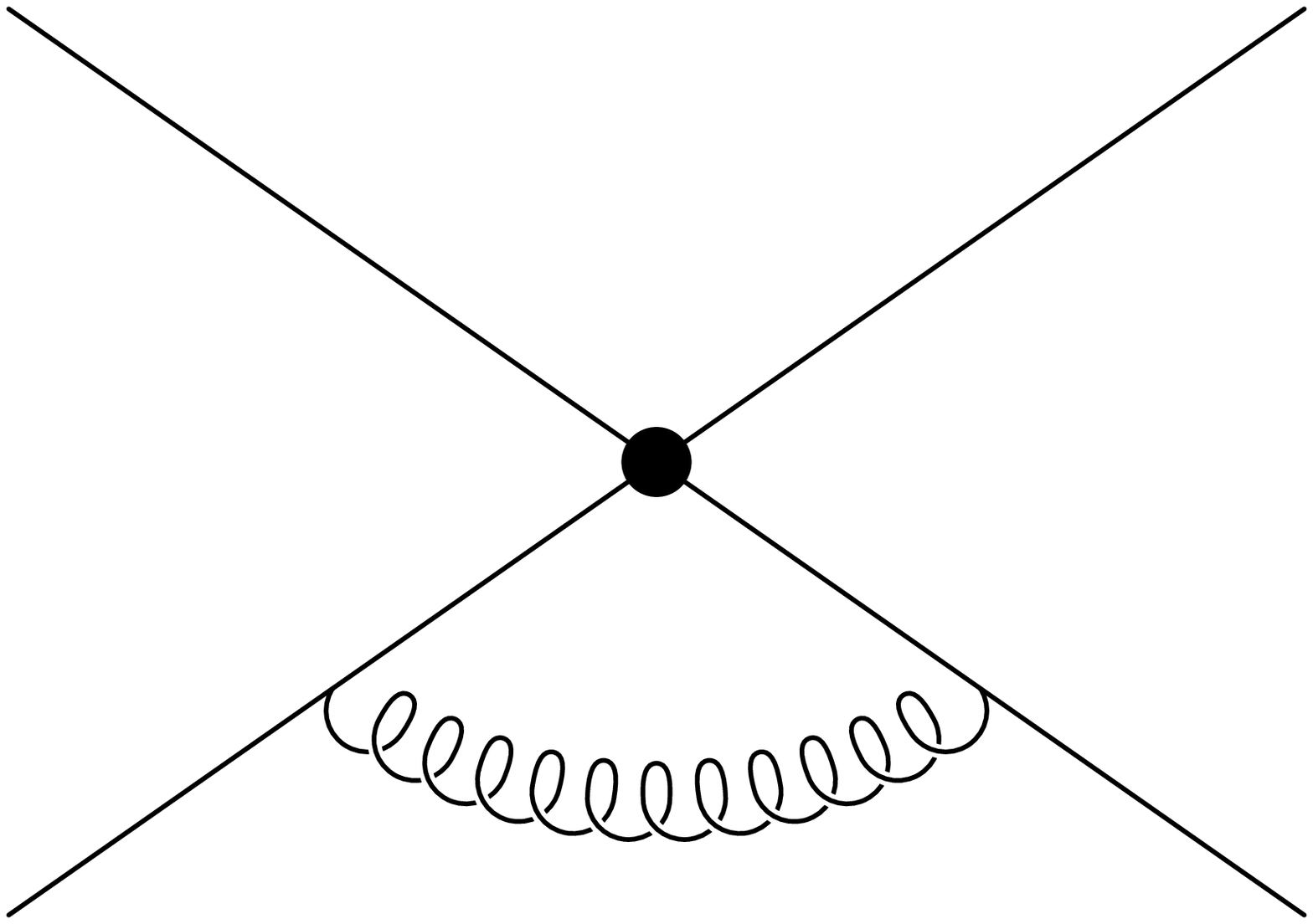}}  & &
 X^a \otimes \bar T^b \bar X^a \bar T^b & {\mathbf p \cdot p^\prime} &
 -{1\over 3}\left({\mathbf p^2}+{\mathbf p^{\prime\,2}}+
 {\mathbf p \cdot p^\prime}\right) \\ \\
\epsfxsize=2cm \lower15pt\hbox{\epsfbox{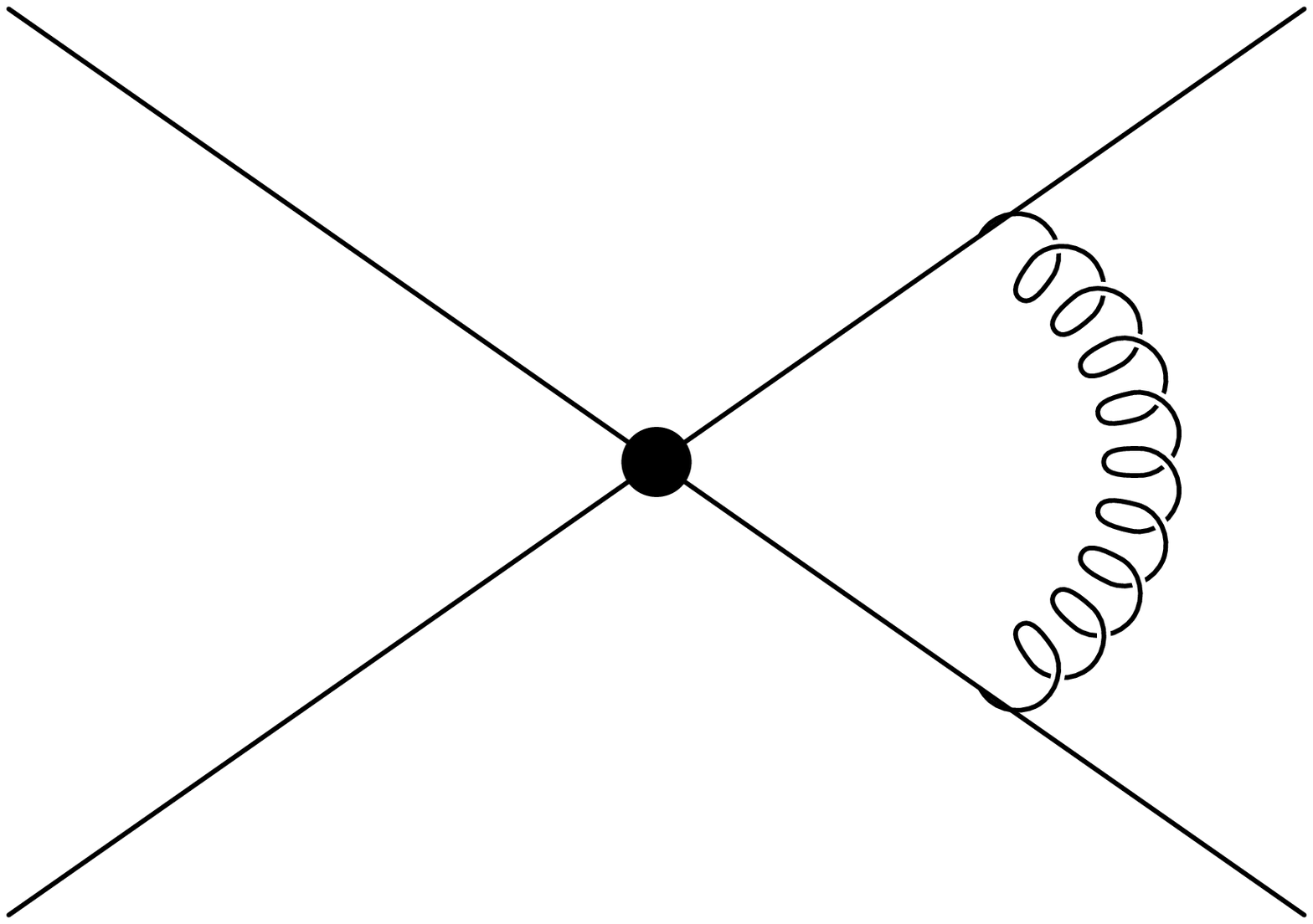}}  & &
 T^b X^a \otimes \bar T^b \bar X^a & {\mathbf p^{\prime\,2}} &
 {1\over 3}{\mathbf p^{\prime\,2}} \\ \\
\epsfxsize=2cm \lower15pt\hbox{\epsfbox{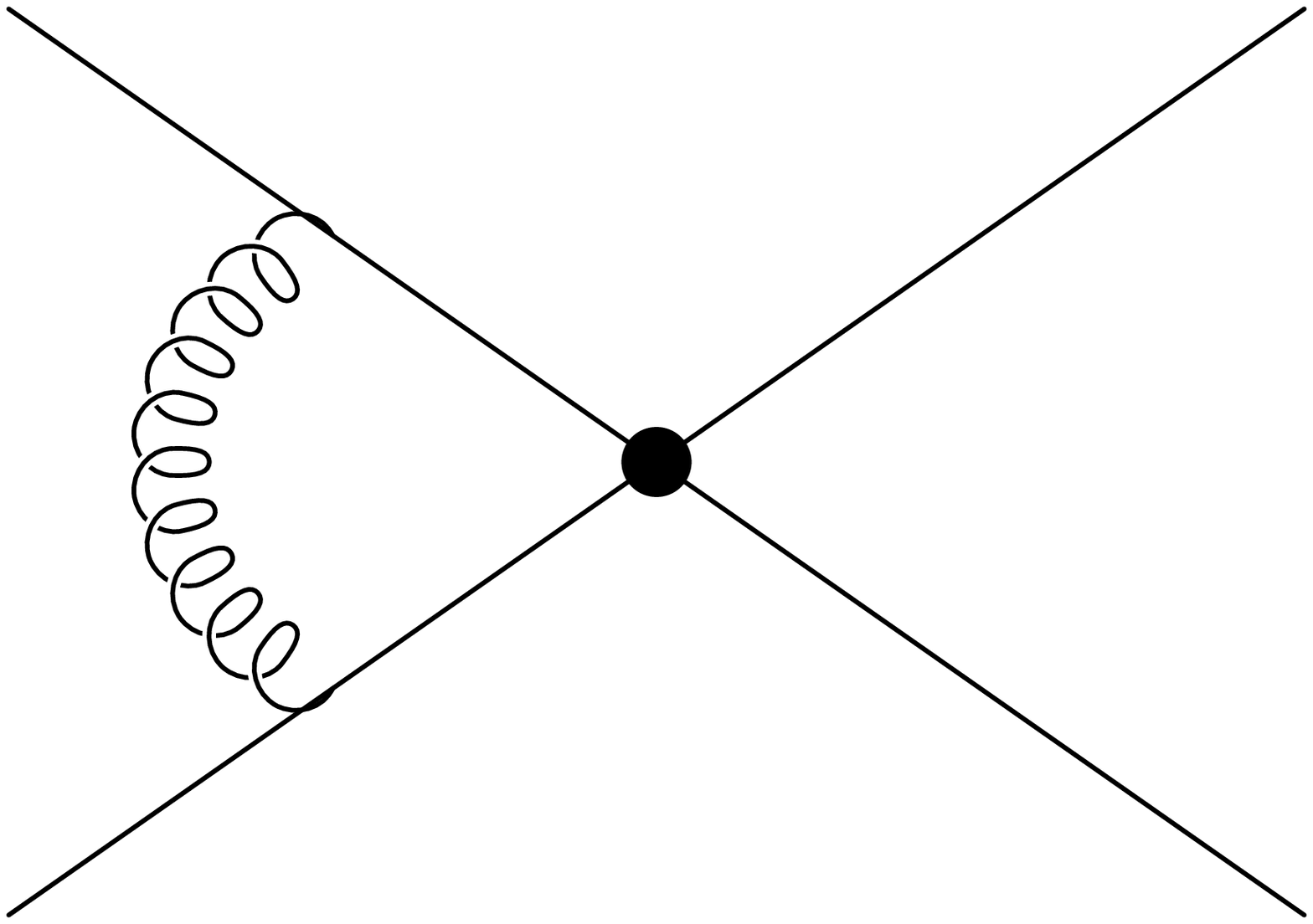}} & &
 X^a T^b \otimes \bar X^a \bar T^b & {\mathbf p^2} &
 {1\over 3}{\mathbf p^2} \\ \\
\epsfxsize=2cm \lower15pt\hbox{\epsfbox{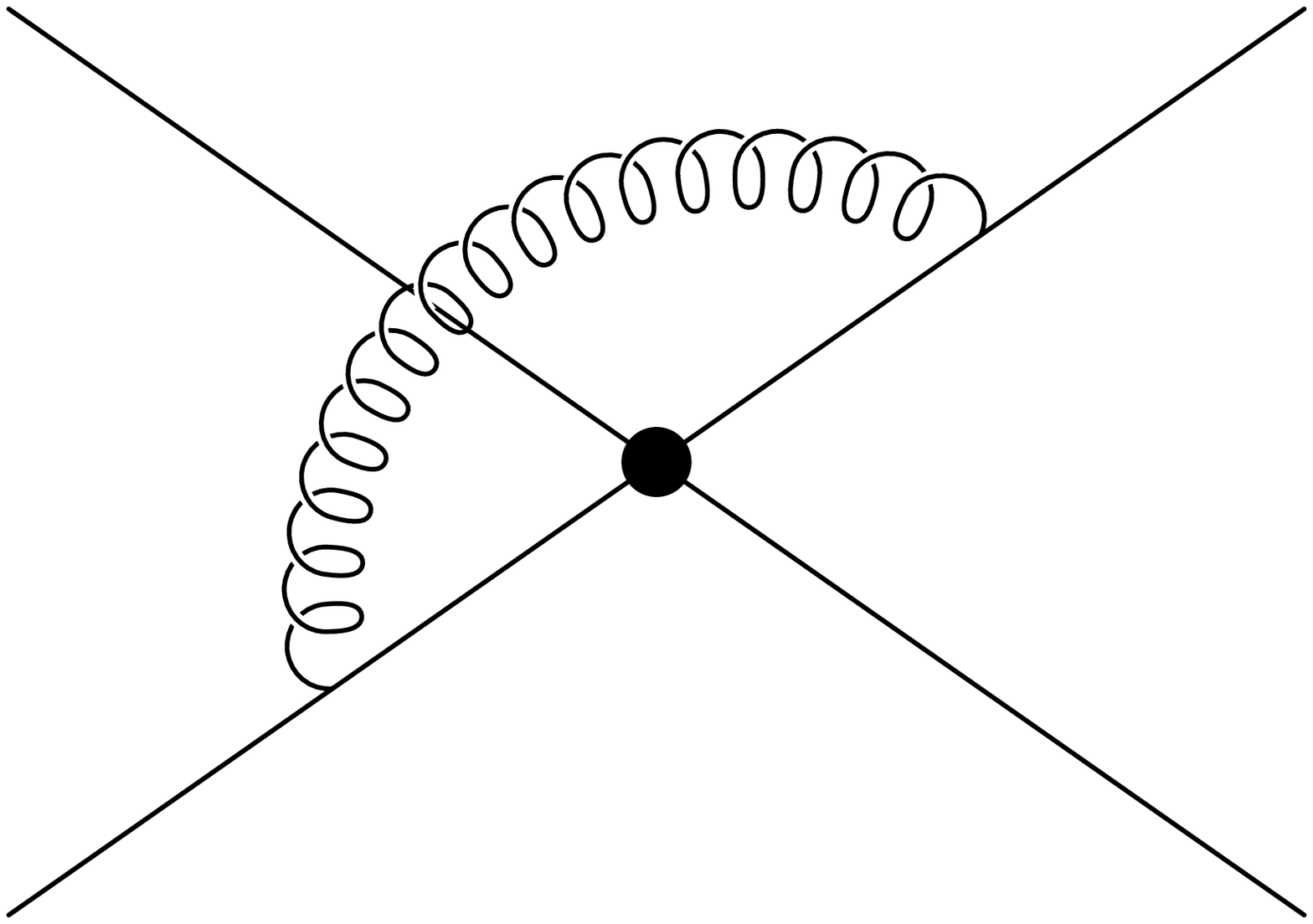}}  & &
 T^b X^a \otimes \bar X^a \bar T^b & -{\mathbf p \cdot p^\prime} &
 -{1\over 3}\left({\mathbf p^2}+{\mathbf p^{\prime\,2}}-
 {\mathbf p \cdot p^\prime}\right)  \\ \\
\epsfxsize=2cm \lower20pt\hbox{\epsfbox{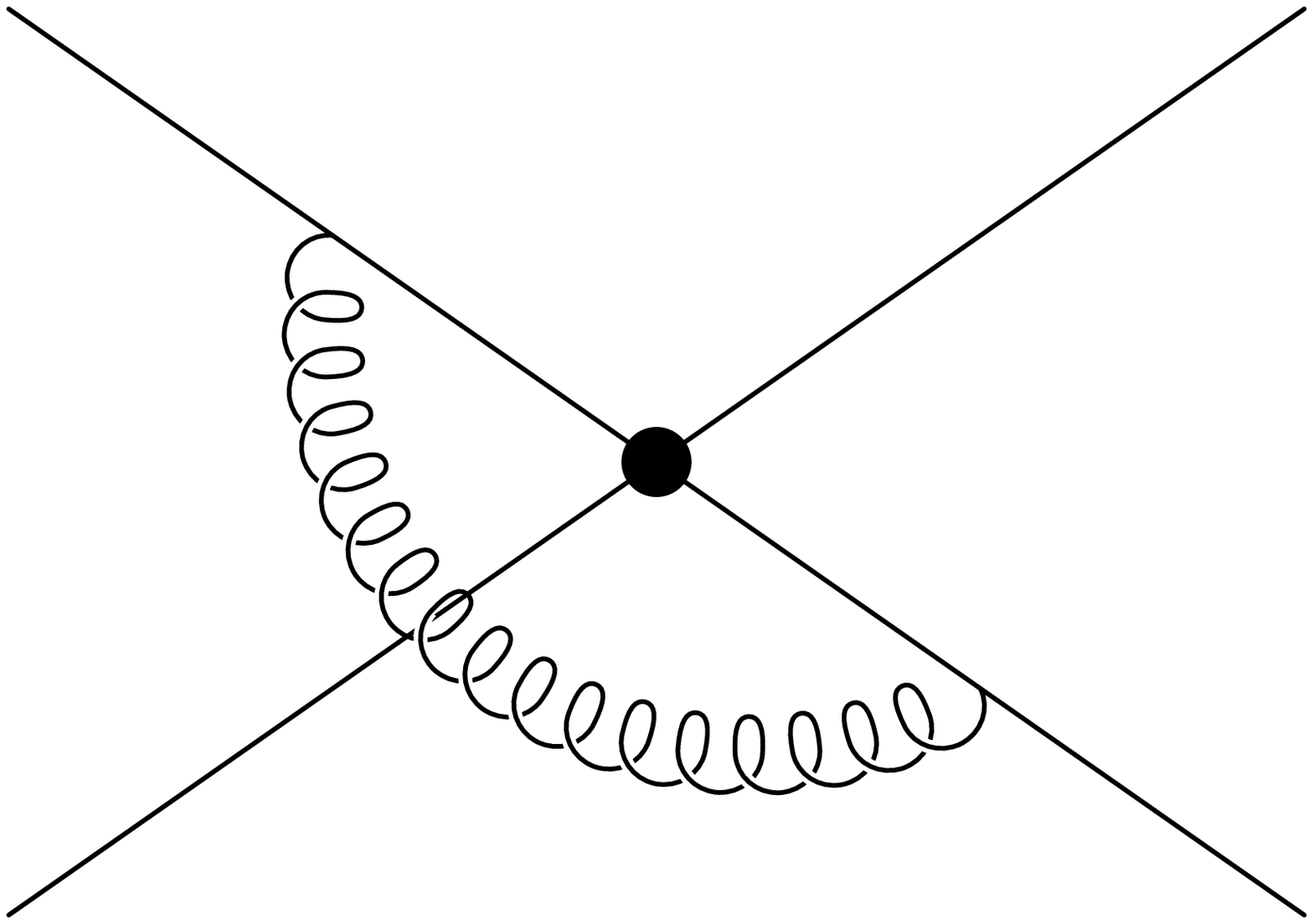}}  & &
 X^a T^b \otimes  \bar T^b \bar X^a & -{\mathbf p \cdot p^\prime} &
 -{1\over 3}\left({\mathbf p^2}+{\mathbf p^{\prime\,2}}-
 {\mathbf p \cdot p^\prime}\right) \\ \\
 \hline\\
\epsfxsize=2cm \lower20pt\hbox{\epsfbox{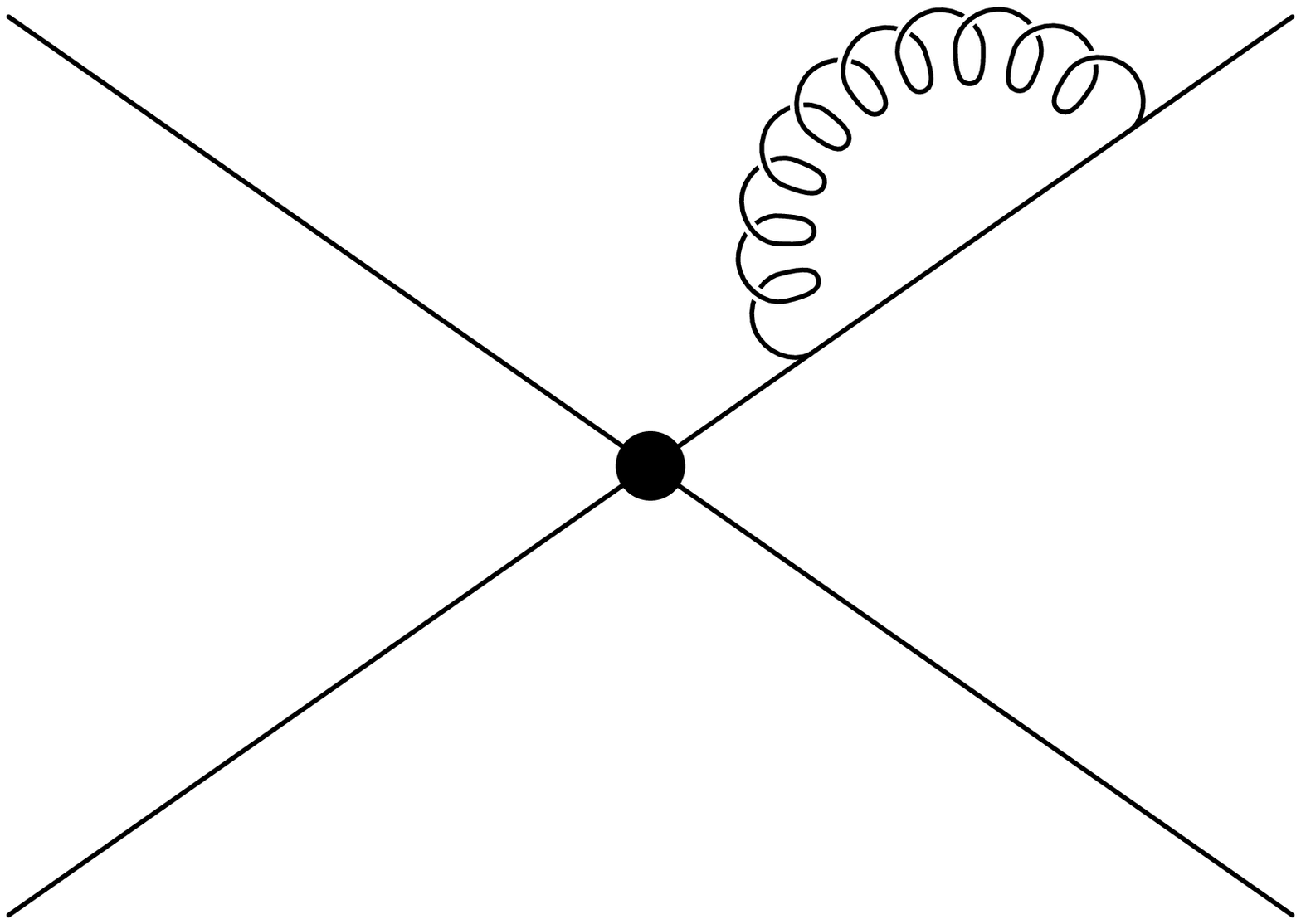}} &  &
 C_F X^a \otimes \bar X^a & - \left( {\mathbf p^2 + p^{\prime\,2}} \right) &
 {\mathbf p^2 + p^{\prime\,2}} \\
\end{array}
\]
\end{table}
\begin{eqnarray}  \label{Gusoft}
  \Gamma^{(us)} = { 2 \over 3 m^2 }\left[\begin{array}{ccc}
  C_F {\mathbf k^2} & \phantom{xx} & - C_1 {\mathbf k^2} \cr
  -{\mathbf k}^2 & \phantom{xx}
  & (C_F+\frac14 C_d -\frac34 C_A) {\mathbf k^2} + C_A ({\mathbf p^2+p'^2})\cr
\end{array}\right] \,,
\end{eqnarray}
where
\begin{equation}
C_1={N_c^2-1 \over 4N_c^2} ={C_F T_F \over {\rm dim}_F},\qquad
d^{ABC} d^{ABD}= C_d\, \delta^{CD},
\end{equation}
so $C_d=(N_c-4/N_c)$. Here ${\rm dim}_F$ is the dimension of the fermion
representation. In the VRG method introduced in Ref.~\cite{LMR}, the scale
$\mu$ in ultrasoft loops is $m\nu^2$, and the ultrasoft coupling constant is
$\alpha_s(m \nu^2)$. Therefore, the ultrasoft contribution to Eq.~(\ref{3}) can
be written as
\begin{eqnarray}
 &&\nu {d \over d \nu} \left[\begin{array}{c} V_{1\otimes 1} \cr V_{T\otimes T}
 \end{array}\right] = { 2 \alpha_{s}(m\nu^2) \over \pi} \: \Gamma^{(us)} \:
 \left[\begin{array}{c} V_{1\otimes 1} \cr V_{T\otimes T} \end{array}\right]
 \,. \label{ADusoft}
\end{eqnarray}
From Eqs.~(\ref{Gusoft}) and (\ref{ADusoft}) we see that the Coulomb potential
induces running in ${\cal V}_2^{(1)}$, ${\cal V}_2^{(T)}$, and ${\cal
V}_r^{(T)}$ proportional to ${\cal V}_c^{(T)}$.

\begin{figure}
  \epsfxsize=4cm \centerline{\epsfbox{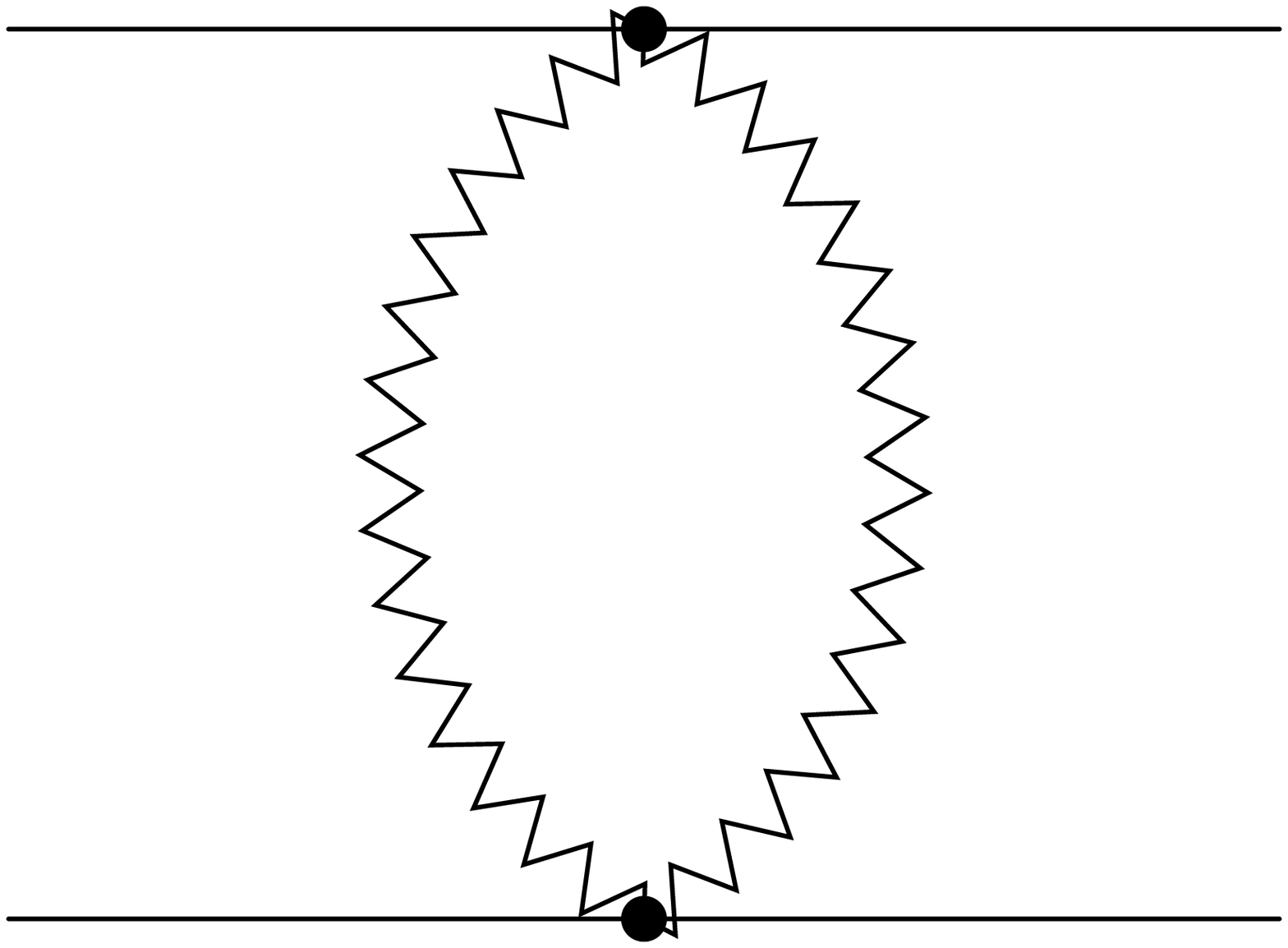}}
  \medskip
{\tighten \caption{Soft contributions to the running of the potential. The loop
includes soft gluons, ghosts, and massless quarks. Graphs with two $\sigma=0$
vertices from Eq.~(\ref{Lsoft}) gives the running of the coefficient ${\cal
V}_c^{(T)}$. Graphs with one $\sigma=0$ and one $\sigma=1$ vertex vanish, while
those with one $\sigma=0$  and one $\sigma=2$, or two $\sigma=1$ vertices
contribute to the running of the order $v^2$ corrections to the potential.}
\label{fig_sloop} }
\end{figure}

In addition to the ultrasoft loops, one has the soft contribution shown in
Fig.~\ref{fig_sloop}.  For the soft gluon loops all the infrared divergences
are converted to ultraviolet divergences by tadpole diagrams, so with
dimensional regularization all $1/\epsilon$ poles contribute to the anomalous
dimensions.
The divergent parts of the soft gluon, ghost, and quark loops in
Fig.~\ref{fig_sloop} give the running
\begin{eqnarray}
\mu {d \over d \mu} \left[\begin{array}{c} V_{1\otimes 1} \cr V_{T\otimes T}
\end{array}\right] &=& \alpha_s^2 \: \left[\begin{array}{c}
  B_{1\otimes 1} \cr B_{T\otimes T}  \end{array}\right] \,,
\end{eqnarray}
where for $n_f$ massless quarks we find
\begin{eqnarray}
  B_{1\otimes 1} &=& { {14 \over 3 m^2}}\, C_1  \,,\\[5pt]
  B_{T\otimes T} &=& -  {7 \over 6 m^2} C_d + C_A \Bigg\{
  - {22 \over 3 {\mathbf k}^2}+ {(39 + 4 c_F^2) \over 6 m^2} -
   {19({\mathbf p^2+p^{\prime 2})} \over 3 m^2 {\mathbf k}^2 }
   +  {(11 c_S+10 c_F) \over 3 m^2} \Lambda({\mathbf p',p})  \nn\\*
  & & + { c_F^2 \over 9 m^2} {\mathbf S}^2 +
  {5 c_F^2 \over 18 m^2} T({\mathbf k}) \Bigg\} + \frac{8 T_F n_f}{3}
  \Bigg\{   {1 \over {\mathbf k}^2} + { (2 c_F^2-c_D-1) \over 4 m^2 }
  + {({\mathbf p^2+p^{\prime 2})} \over 2 m^2 {\mathbf k}^2 } \nn\\*
  &&  -  {(c_S+2 c_F) \over 2 m^2} \Lambda({\mathbf p',p}) -
   { c_F^2 \over 3 m^2} {\mathbf S}^2-  {c_F^2 \over 12 m^2} T({\mathbf k})
   \Bigg\} \,. \nn
\end{eqnarray}
Here $B_{1\otimes 1}$ and $B_{T\otimes T}$ depend on the scale $\mu=m\nu$
through their dependence on $c_F$, $c_D$, and $c_S$.  In the VRG, the soft
coupling constant is $\alpha_s(m\nu)$, so the soft contribution to the running
is
\begin{eqnarray}  \label{ADsoft}
 \nu {d \over d \nu} \left[\begin{array}{c} V_{1\otimes 1} \cr V_{T\otimes T}
  \end{array}\right]&=& \alpha_s^2(m\nu)\ \left[\begin{array}{c}
  B_{1\otimes 1}(m\nu) \cr B_{T\otimes T}(m\nu)  \end{array}\right]  \,.
\end{eqnarray}
The total VRG running of the potential is given by adding Eqs.~(\ref{ADusoft})
and (\ref{ADsoft}).  The running of the Coulomb potential is proportional to
the beta function,
\begin{eqnarray} \label{vcrun}
  \nu {\partial \over \partial\nu} {\cal V}_c^{(T)} = -2 \beta_0\,
  \alpha_s^2(m\nu)  \,.
\end{eqnarray}
Integrating Eq.~(\ref{vcrun}) and using the boundary condition in
Eq.~(\ref{bc}) gives the solution
\begin{eqnarray} \label{vcsoln}
  {\cal V}_c(\nu) = 4\pi \alpha_s(m\nu) \,,
\end{eqnarray}
and large logarithms can be avoided by choosing $\nu=|{\mathbf k}|/m$. This
gives the expected result that the renormalization group improved Coulomb
potential is given by choosing $\alpha_s=\alpha_s(|{\mathbf k}|)$.

Order $v$ corrections to the potential such as the one in Eq.~(\ref{Lk}) do
not run at one-loop. The non-trivial result is the renormalization group
equations for the order $v^2$ terms in the potential.
Combining Eqs.~(\ref{ADusoft}), (\ref{ADsoft}), and using (\ref{vcsoln}) we 
find
\newpage
\begin{eqnarray}  \label{VRG}
\nu {\partial \over \partial\nu} {\cal V}_r^{(T)} &=& -2 (\beta_0 + \frac{8}{3}
  C_A)\, \alpha_s^2(m\nu) + \frac{32}{3} C_A\, \alpha_s(m\nu) \alpha_s(m\nu^2)
  \,, \nn\\[5pt]
\nu {\partial \over \partial\nu} {\cal V}_2^{(T)} &=&
  \left\{ \frac12 \beta_0 \Big[1+c_D(\nu)-2 c_F^2(\nu) \Big]+
  \frac{1}{6} C_A \Big[28 - 11 c_D(\nu) + 26 c_F(\nu)^2 \Big] -
  \frac{7}{6} C_d \right\} \, \alpha_s^2(m\nu) \nn\\
 &&  + \frac{4}{3} (4C_F+C_d-3C_A)\, \alpha_s(m\nu) \alpha_s(m\nu^2) \,,
 \nn\\[5pt]
\nu {\partial \over \partial\nu} {\cal V}_2^{(1)} &=& \frac{14}{3}\, C_1\,
  \alpha_s^2(m\nu) - \frac{16}{3}\,C_1\, \alpha_s(m\nu) \alpha_s(m\nu^2) \,,
  \nn \\[5pt]
\nu {\partial \over \partial\nu} {\cal V}_s^{(T)} &=& \frac13 (2\beta_0-7 C_A)
  \, c_F^2(\nu)\: \alpha_s^2(m\nu) \,, \nn \\[5pt]
\nu {\partial \over \partial\nu} {\cal V}_t^{(T)} &=& \frac16 (\beta_0-2 C_A)
 \, c_F^2(\nu)\: \alpha_s^2(m\nu) \,, \nn \\[5pt]
\nu {\partial \over \partial\nu} {\cal V}_\Lambda^{(T)} &=& \Big\{
  \beta_0 [c_S(\nu)+2c_F(\nu)]-4 C_A c_F(\nu) \Big\} \,\alpha_s^2(m\nu) \,.
\end{eqnarray}
Note that the soft contributions to the running depend on $\alpha_s^2(m\nu)$,
and the ultrasoft contributions to the running depend on $\alpha_s(m\nu)
\alpha_s(m\nu^2)$. This is because the soft gluon coupling is $g(m\nu)$, and
the ultrasoft gluon coupling is $g(m\nu^2)$. The ultrasoft gluon couples via a
multipole-expanded interaction,  so the ultrasoft interaction vertex does not
involve momentum transfers of order $mv$.

Our results can be checked by comparing with the one-loop formula for the color
singlet quark potential in Ref.~\cite{Yndurain}.  The nonrelativistic expansion
of the QCD calculation has $\ln(|{\mathbf k}|/m)$, $\ln(\mu/m)$, and
$\ln(\lambda/m)$ terms, where a finite gluon mass $\lambda$ was introduced as
an infrared regulator. By explicit computation of the box and crossed box
diagrams \cite{Redhead,Nieuw} we found that the spin independent potential in
Eq.~(19) of Ref.~\cite{Yndurain} is missing the order $v^2$ term\footnote{Note
that in the calculation in Ref.~\cite{Gupta} an expansion was made
in $({\mathbf
p+p'})^2/{\mathbf k}^2$, so that the term in Eq.~(\ref{dropped})
was dropped.}
\begin{eqnarray}\label{dropped}
  -\frac43\, C_F C_A\, \alpha_s^2\, \frac{({\mathbf p+p'})^2}{m^2{\mathbf k}^2}
   \, \ln\left(\frac{\lambda}{|\mathbf k|}\right) \,.
\end{eqnarray}
The $\ln(|{\mathbf k}|)$ and $\ln(\lambda)$ dependence in the full theory
should be reproduced by the effective theory, so with $\mu=m$ all QCD logs
should be reproduced. We find that the $\log(\lambda/m)$ terms are reproduced
by the ultrasoft diagrams in Table~\ref{tab:us} and that the $\log(|{\mathbf
k}|/m)$ terms are reproduced by the soft loops in Fig.~\ref{fig_sloop}.

Solving the VRG equations in Eq.~(\ref{VRG}) with the tree level boundary
conditions in Eq.~(\ref{bc}) gives\footnote{In NRQED the results
are simpler since the coupling in the effective theory does not run.}
\newpage
\begin{eqnarray}  \label{pv2}
 {\cal V}_r^{(T)}(\nu) &=& 4\pi\,\alpha_s(m\nu) - \frac{16 C_A}{3}\,
  \alpha_s(m\nu)\alpha_s(m) \ln\Big({m\nu\over m}\Big)
   + \frac{64\pi C_A}{3\beta_0}\,
   \alpha_s(m)\ln\bigg[ \frac{\alpha_s(m\nu)}{\alpha_s(m\nu^2)} \bigg] \,,
   \nn \\
 {\cal V}_2^{(T)}(\nu) &=& \frac{\pi\Big[ C_A (352 C_F+91 C_d-144 C_A)
    - 3\beta_0 (33 C_A+32 C_F) \Big]}{39\beta_0 C_A} \
    \Big[\alpha_s(m\nu) - \alpha_s(m) \Big] \nn\\
  &+& \frac{8\pi (3\beta_0-11C_A)(5 C_A+8 C_F) \alpha_s(m)}
    {13 C_A (6\beta_0-13 C_A)} \Big[ z^{(1-13 C_A/(6\beta_0))} - 1 \Big]
    \nn \\
  &+& \frac{ \pi (\beta_0-5 C_A)\alpha_s(m)}{(\beta_0-2 C_A)}
   \Big[ z^{(1-2C_A/\beta_0)} - 1 \Big]
    + \frac{8\pi (4C_F\!+\!C_d\!-\!3C_A)}{3\beta_0}
    \, \alpha_s(m) \ln\bigg[ \frac{\alpha_s(m\nu)}{\alpha_s(m\nu^2)} \bigg]
    \,, \nn \\
 {\cal V}_2^{(1)}(\nu) &=& \frac{14 C_1}{3}\,
   \alpha_s(m\nu)\alpha_s(m) \ln\Big({m\nu\over m}\Big) -
    \frac{32\pi C_1}{3\beta_0}
    \, \alpha_s(m) \ln\bigg[ \frac{\alpha_s(m\nu)}{\alpha_s(m\nu^2)} \bigg]
    \,, \nn \\
 {\cal V}_s^{(T)}(\nu) &=& \frac{2 \pi \alpha_s(m)}{(2 C_A-\beta_0) } \bigg[
   C_A + \frac{1}{3} ( 2\beta_0 - 7 C_A) \ z^{(1-2 C_A/\beta_0)} \bigg]
    \,, \nn\\
 {\cal V}_s^{(1)}(\nu) &=& 0 \,, \nn \\
 {\cal V}_t^{(T)}(\nu) &=& -\frac{\pi\alpha_s(m)}{3}\,
    \ z^{(1-2 C_A/\beta_0)}  \,,\nn \\
 {\cal V}_\Lambda^{(T)}(\nu) &=& 2 \pi\alpha_s(m) \Big[ z - 4\
    z^{(1-C_A/\beta_0)}  \Big]  \,,
\end{eqnarray}
where $z=\alpha_s(m\nu)/\alpha(m)$. The annihilation contributions can be
accounted for by adding the expressions in Eq.~(\ref{bc2}) to Eq.~(\ref{pv2}).

The logarithmic corrections to the coefficients in the quark potential were
considered by Brambilla et al.~\cite{Brambilla} using pNRQCD, but
were not resummed. Since only the leading perturbative
logarithms were included, we can compare our results to theirs by expanding the
resummed logarithms in Eq.~(\ref{pv2}).  Brambilla et al.\ give expressions for
the ${\cal V}$'s in the color singlet channel containing terms of the form
\begin{eqnarray}
  \alpha_s(r),\qquad  \alpha_s\,\alpha_s(r)\ln(m r),\qquad
  \alpha_s\,\alpha_s(r)\ln(\mu r) \,.
\end{eqnarray}
For the soft contributions letting $\alpha_s(r)\to \alpha_s(m\nu)$ and
$\ln(r)\to \ln(1/m\nu)$ in Ref.~\cite{Brambilla} gives agreement\footnote{When
comparing the expansion of the color singlet combination ${\cal V}_2^{(1)}-C_F
{\cal V}_2^{(T)}$ the replacements $C_d\to 8C_F-3C_A$ and $C_1\to C_F
C_A/2-C_F^2$ are also necessary to obtain agreement.} with expanding the soft
contributions in Eq.~(\ref{pv2}) about $z=1$ (recalling that
$z=1-\beta_0/(2\pi) \alpha_s(m\nu) \ln(m\nu/m)$). For the ultrasoft
contributions letting $\ln(\mu/m)\to \ln(m\nu^2/m)$ in Ref.~\cite{Brambilla}
gives agreement with the ultrasoft contributions in Eq.~(\ref{pv2}) with the
expansion
\begin{eqnarray}
 \ln\bigg[ \frac{\alpha_s(m\nu)}{\alpha_s(m\nu^2)} \bigg] =
 \frac{\beta_0}{2\pi}\, \alpha(m\nu) \ln\Big(\frac{m\nu^2}{m\nu}\Big)+\ldots =
 \frac{\beta_0}{4\pi}\, \alpha(m\nu) \ln\Big(\frac{m\nu^2}{m}\Big)+\ldots \,.
\end{eqnarray}
It would be interesting to compare our resummed coefficients to the analogous
results computed in pNRQCD. Recall that there is an important distinction between
pNRQCD and vNRQCD. In pNRQCD the scales $mv$ and $mv^2$ are treated
independently, while vNRQCD builds in the fact that the scales $mv$ and $mv^2$
are not independent. Thus, in vNRQCD, the scaling of soft terms from $m$ to $mv$,
and ultrasoft terms from $m$ to $mv^2$, occurs simultaneously, and $\ln(mv)$ and
$\ln(mv^2)$ terms are resummed together.

To see the effect of the running on the value of the coefficients in the
potential, consider the case of top quark production near threshold. Projecting
onto the color singlet channel gives the singlet coefficients ${\cal
V}^{(s)}_i={\cal V}^{(1)}_i- C_F {\cal V}^{(T)}_i$. 
Using $\alpha_s(m_t)=0.108$, the tree level values in Eq.~(\ref{bc}) are:
\begin{eqnarray}
 {\cal V}_r^{(s)} = -1.81 \,, \quad
 {\cal V}_2^{(s)} = 0 \,, \quad
 {\cal V}_s^{(s)} = 0.60 \,, \quad
 {\cal V}_\Lambda^{(s)} = 0.15 \,,\quad
 {\cal V}_t^{(s)} = 2.71 \,.
\end{eqnarray}
For a Coulombic system the velocity $v$ is determined by solving
$\alpha_s(mv)=v$.  Using $m_t=175\,{\rm GeV}$ and the
one-loop running of $\alpha_s(\mu)$ with $n_f=5$ gives $v=0.145$. For $\nu=v$
the running coefficients in Eq.~(\ref{pv2}) are:
\begin{eqnarray}
 {\cal V}_r^{(s)} = -1.49 \,, \quad
 {\cal V}_2^{(s)} = 0.63 \,, \quad
 {\cal V}_s^{(s)} = 0.53 \,, \quad
 {\cal V}_\Lambda^{(s)} = 0.16 \,,\quad
 {\cal V}_t^{(s)} = 3.11 \,.  \label{res1}
\end{eqnarray}
The most substantial change is to the value of ${\cal V}_2^{(s)}$ which was
zero at tree level.  Using the results of Brambilla et al.\ with $m r=1/v$ and
$\mu=mv^2$ gives
\begin{eqnarray}
 {\cal V}_r^{(s)} = -1.78 \,, \quad
 {\cal V}_2^{(s)} = 0.68 \,, \quad
 {\cal V}_s^{(s)} = 0.53 \,, \quad
 {\cal V}_\Lambda^{(s)} = 0.16 \,,\quad
 {\cal V}_t^{(s)} = 3.15 \,, \label{res2}
\end{eqnarray}
indicating that resummation of the logarithms has the largest effect on ${\cal
V}_r^{(s)}$.  For the remaining coefficients the resummed values in
Eq.~(\ref{res1}) are fairly close to the coefficients with perturbative
logarithms in Eq.~(\ref{res2}).

We would like to thank M.~Luke, A.~Pineda, and I.~Rothstein for helpful
discussions.  This work was supported in part by the Department of Energy under
grant DOE-FG03-97ER40546.

{\tighten

} 

\end{document}